\documentclass[aps,prl,twocolumn,groupedaddress,showpacs, floatfix]{revtex4}

\usepackage{graphicx}

\begin{document}


\title{Observation of inter-edge magnetoplasmon mode in a degenerate
two-dimensional electron gas}


\author{G. Sukhodub}
\email[Corresponding author:
]{sukhodub@nano.uni-hannover.de}
\author{F. Hohls}
\author{R. J. Haug}
\affiliation{Institut f\"ur Festk\"orperphysik,
Universit\"at Hannover, Appelstr. 2, D-30167 Hannover,
Germany}


\date{\today}

\begin{abstract}
We study the propagation of edge magnetoplasmons by
time-resolved current measurements in a sample which allows
for selective detection of edge states in the quantum Hall
regime. We observe two decoupled modes of edge and
inter-edge magnetoplasmons at filling factors close to 3.
From the analysis of the propagation velocities of each mode
the internal spatial parameters of the edge structure are
derived.
\end{abstract}

\pacs{73.43.Lp, 73.20.Mf}

\maketitle



%

Low-frequency collective excitations of the charge density
in a two-dimensional electron system (2DES) in a magnetic
field travel along the edge of the sample and are called
edge magnetoplasmons (EMP's). Since their first observation
by \mbox{Allen \textit{et. al.} \cite{allen83}} EMP's have
been extensively studied both theoretically
\cite{volk88,balev,emp_book} and experimentally
\cite{emp_book,talyan92,zhit93}. An inter-edge
magnetoplasmon (IEMP) mode \cite{mikhailov92,mikhailov95}
attributed to the excitations of the boundary between two
regions with constant but different charge densities was
experimentally observed up to now only in a classical system
of surface electrons on helium
\cite{sommerfeld95,kirichek95}.

In a degenerate 2DES under quantum Hall effect conditions
different density profiles separated by incompressible
strips with constant charge densities appear \cite{chkl92}.
The influence of edge channels on the propagation of EMP's
was addressed in experiments on screened 2DES's in the
frequency domain \cite{talyan92,talyan95} and in the time
domain \cite{zhit93,zhit95}. The splitting of an incident
voltage pulse into several EMP modes propagating with
different velocities was observed in unscreened samples
\cite{ernst96}. The additional slower modes were interpreted
as acoustic EMP modes \cite{aleiner94}. In none of these
experiments with non-classical systems IEMP modes were
identified.

\begin{figure}
\includegraphics{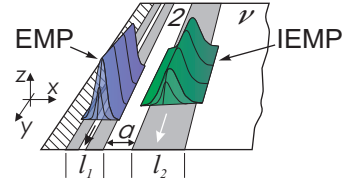}
\caption{\label{fig:sketch} Schematic view of EMP and IEMP
modes propagating with different velocities and decoupled by
an incompressible strip with filling factor $\nu=2$.
Compressible strips are shown as grey areas, incompressible
liquids as white areas, and the hatched region represents a
depleted semiconductor at the mesa edge.}
\end{figure}

We report here the observation of both EMP and IEMP modes in
a screened 2DES in the quantum Hall regime. In our
experiment we clearly demonstrate the decomposition of the
two-stage rise of the observed current pulse in the vicinity
of the bulk filling factor $\nu=3$ into different modes of
edge magnetoplasmons. Using selective detection of edge
states we can distinguish two wave packets of edge
excitations moving with different velocities near the sample
boundary (Fig.~\ref{fig:sketch}).
The faster one propagates along the compressible region
formed by two strongly coupled outer compressible strips
originating from the lowest Landau level. It gives rise to a
fundamental EMP mode. The slower wave packet propagates
along the inner compressible strip arising from the second
Landau level. This wave packet corresponds to an IEMP mode
\cite{mikhailov92,mikhailov95}.

\begin{figure}
    \includegraphics{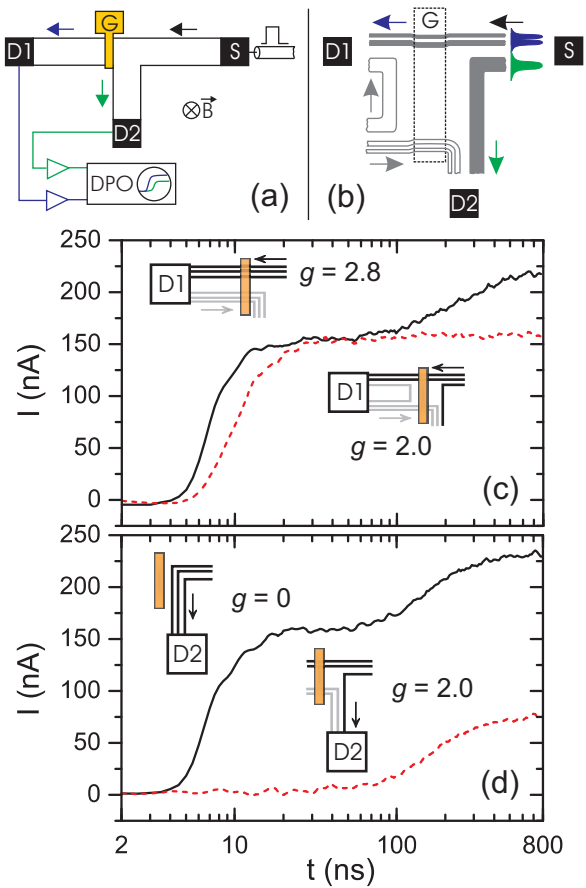}
\caption{\label{fig:seldet}(a) Sketch of experimental setup.
The metallic top gate which covers the whole area of the
sample is omitted for clarity. (b) Scheme of selective
detection of edge states. The local filling factor $g$ under
the gate G is tuned to $g=2$. The populated compressible
strips are shown as dark areas. (c),(d) Transient currents
at $B=2.70$\,T $(\nu=2.8)$ measured at contacts D1 and D2
for the entire edge structure (solid traces in both panels),
the two outer EC's (dashed trace in (c)), and the inner EC
(dashed trace in (d)). Sketches along each curve show the
configuration of detected edge states.}
\end{figure}

A schematic view of the experimental setup is shown in
Fig.~\ref{fig:seldet}(a).
An AlGaAs/GaAs heterostructure is patterned into a T-shaped
form with a width of $200\,\mu$m. The 2DES has a carrier
concentration $n_s=1.8\times10^{15}$\,m$^{-2}$ and a
mobility $\mu=70$\,m$^2$/Vs. A long voltage pulse of
$800$\,ns and typically $2.0$\,mV amplitude with $\sim1$\,ns
rise time is applied to the source contact S. With the
perpendicular magnetic field pointing into the page it
propagates along the upper boundary of the sample to the
drain contact D1. A Schottky gate G crosses the entire
sample and is situated between S and D1. If a negative
voltage is applied to this gate some fraction of the signal
is deflected to the other drain contact D2. The distances
between source S and each of the drain contacts D1 and D2
are equal and amount to $L=1.56$\,mm. Thus, the transient
currents detected at both contacts are directly comparable.
In addition, a metallic top gate is deposited on top of the
entire sample. The distance $d$ between the top gate and the
2DES is $105$\,nm. A thin layer of polymethylmethacrylate
covering the Schottky gate G isolates the gates from each
other in the overlap region. Transmitted pulses are
amplified and recorded by a multi-channel digital phosphor
oscilloscope (DPO). In such a setup the voltage drop at the
input resistance of the preamplifier is proportional to the
current through the drain contact.

The scheme of selective detection of EC is depicted in
Fig.~\ref{fig:seldet}(b). A bulk filling factor $\nu$ is set
by the external magnetic field. Here we concentrate on
$\nu\approx3$ since this choice yields the clearest
occurrence of the IEMP. The contribution of the entire edge
structure can be measured at both drain contacts D1 and D2.
In the first case the gate G remains unbiased
($V_{G}=0$\,mV) and all EC's are transmitted to D1. The
local filling factor $g$ under the gate is equal to the bulk
value $\nu$ [upper sketch in Fig.~\ref{fig:seldet}(c)]. In
the second case a negative voltage of $V_{G}=-350$\,mV is
applied to the gate which depletes the 2DES underneath
completely ($g=0$) deflecting all EC's to D2 [upper sketch
in Fig.~\ref{fig:seldet}(d)]. To measure the contribution of
each Landau level by different contacts the filling factor
$g=2$ is established under the gate G [this situation is
depicted in Fig.~\ref{fig:seldet}(b)]. In this configuration
the two outer EC's formed by the lowest Landau level are
transmitted to D1 and the innermost EC is deflected to D2.

Figure~\ref{fig:seldet} represents the main finding of our
study, where the decomposition of the two-stage rise of the
pulse into two single modes is demonstrated. The
time-resolved currents for $\nu=2.83$ detected at contacts
D1 and D2 are shown in panels (c) and (d) respectively. The
origin of the time axis $t=0$ corresponds to the application
of the pulse to the source S. The upper solid curves in both
panels represent the signal carried by the entire edge
structure. Though different contact pairs were used to
detect these traces they closely resemble each other. In
both cases the delay time $t_0$, which is the travelling
time of the signal between source and drain, is equal to
\mbox{$t_{0}=4.63 \pm 0.14$\,ns} and the first steep rise
saturates completely after $t=30$\,ns to \mbox{$I_{1}=(2.02
\pm 0.03)(e^2/h)V_{in}$}, where $V_{in}=2.0$\,mV is the
pulse amplitude. This first step is followed by a slower
rise which develops at $t>70$\,ns and saturates to
$I=I_{1}+I_{2}$ with \mbox{$I_{2}=(0.92 \pm
0.04)(e^2/h)V_{in}$} for this filling factor.

The current associated with edge channels originating from
the lowest Landau level only is represented by the lower
dashed curve in Fig.~\ref{fig:seldet}(c). In its initial
part it bears a strong similarity to the upper curve: after
$t=30$\,ns the current saturates to $2(e^2/h)V_{in}$. For
longer times it remains constant as in the case of $\nu=2$.
The second slower rise is completely missing in this trace.
The current pulse arrives at D1 as a single wave packet
without being redistributed over the whole edge structure.
The initial parts of both curves in Fig.~\ref{fig:seldet}(c)
reveal minor differences (note the logarithmic scale of $t$
in the figure). Namely, the dashed curve ($g=2$) is delayed
by $\sim1.0$\,ns and shows a slight change in the slope of
the rising part. The first issue is related to the reduced
charge density under the gate. This leads to a broadening of
the edge channels attributed to the lowest Landau level and
as a consequence to a lower propagation velocity under the
gate. The change in slope is probably due to the disturbance
of the propagation of edge excitations caused by the
negatively biased gate. Nevertheless, the lower curve in
Fig.~\ref{fig:seldet}(c) reflects the main features of the
fast propagating EMP mode.

The lower curve in Fig.~\ref{fig:seldet}(d) measured at
contact D2 shows the contribution of the inner EC attributed
to the upper Landau level. This slower mode has a delay time
of $t_0=64.0$\,ns and saturates to $I_2\approx(e^2/h)V_{in}$
after several hundred nanoseconds. There is no contribution
of the fast mode in this measurement, whereas it resembles
very much the second slower rise in the solid curve in
Fig.~\ref{fig:seldet}(d). Here we have detected the pure
IEMP mode. To our knowledge, this is the first observation
of an IEMP mode in time-resolved current measurements in a
degenerate 2DES.

The observability of decoupled EMP modes depends on the
relation between the widths $a$, $l_1$, and $l_2$ of
incompressible and compressible strips (see sketch in
Fig.~\ref{fig:sketch}) and the distance $d$ from the top
gate to 2DES. If the condition $d \ll a, l_1, l_2$ is
fulfilled, the electric field associated with each mode can
be assumed to be localized within the corresponding
compressible strip and is determined by the gradient of the
charge density across this strip \cite{zhit95}. The width of
an incompressible strip is substantially smaller than the
width of the surrounding compressible ones \cite{chkl92}. It
is roughly proportional to the gap $\Delta E$ in the energy
spectrum of a 2DES in a perpendicular magnetic field. Using
the effective $g$-factor value $|g^{*}|=0.44$ for bulk GaAs
we estimate the width $a^{(1)}$ of the incompressible strip
with filling factor $\nu^{(1)}=1$ and $\Delta
E=g^{*}\mu_{B}B$ to be comparable to the magnetic length.
Thus, we do not expect to observe decoupled transport due to
this strip, and indeed the contribution of the outer two
compressible strips is represented by a single step in our
experiment [lower curve in Fig.~\ref{fig:seldet}(c)]. The
cyclotron gap $\Delta E=\hbar\omega_{c}$, being
significantly larger, gives rise to a broader incompressible
strip with $a^{(2)}>100$\,nm facilitating the observation of
decoupled edge state transport.

\begin{figure}[b]
\includegraphics{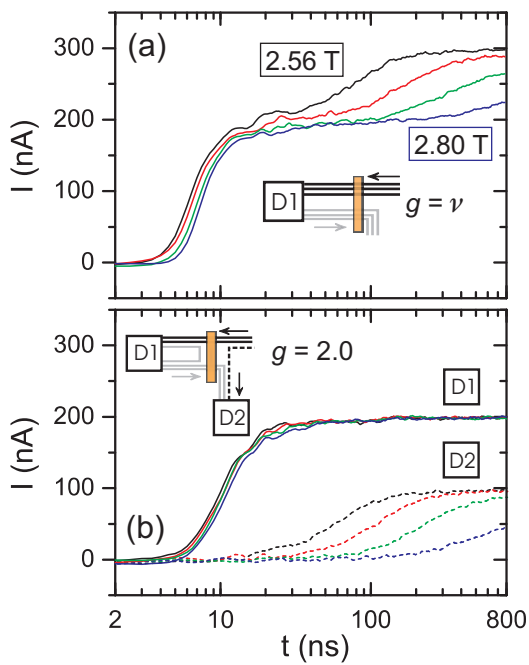}
\caption{\label{fig:magndep}Time-dependent currents measured
at $B=2.56,\,2.64$, $2.72$, and $2.80$\,T
($\nu=2.99,\,2.90,\,2.81$, and $2.73$) for the entire edge
structure (a) and for the contributions of the outer and
inner compressible regions attributed to EMP and IEMP modes
(b). The input pulse voltage is $V_{in}=2.5$\,mV.}
\end{figure}

\begin{figure}
\includegraphics{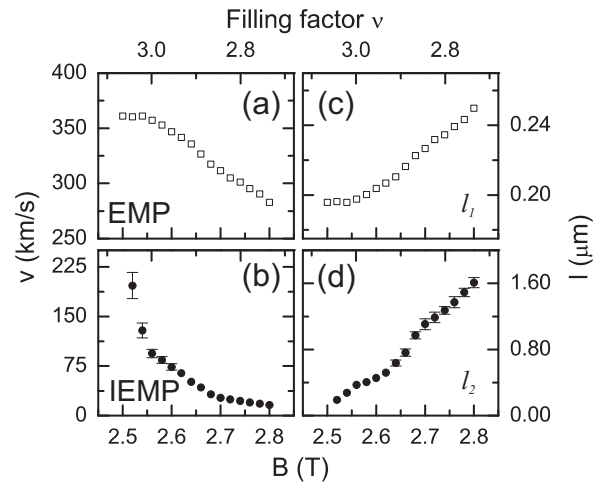}
\caption{\label{fig:vel} Velocities of edge magnetoplasmons
(a) and inter-edge magnetoplasmons (b). Right part shows
calculated widths of compressible regions arising from the
first (c) and the second (d) Landau levels according to
Eq.~(\ref{eqn:v}).}
\end{figure}

Current traces at different magnetic fields around $\nu=3$
are shown in Fig.~\ref{fig:magndep} for both modes measured
together [Fig.~\ref{fig:magndep}(a)] and for each mode
measured independently [Fig.~\ref{fig:magndep}(b)].
We clearly observe a decrease in the velocity with magnetic
field for both modes, although this is much more pronounced
for the slower mode. It is remarkable that the fast mode
[Fig.~\ref{fig:magndep}(b), D1] does not change its form
considerably in the filling factor range $3>\nu>2$ where its
amplitude is ruled by $2e^2/h$. For a quantitative
description of the velocities of the fast mode we analyzed
the initial parts ($t<70$\,ns) of the current traces
corresponding to the entire edge structure
[Fig.~\ref{fig:magndep}(a)]. The IEMP mode velocities were
derived from signals detected at D2 for the case $g=2$
[lower set of curves in Fig.~\ref{fig:magndep}(b)]. The
delay time $t_0$ was obtained from the fit with an
exponential decay function \cite{sukhodub04}. The velocities
of EMP and IEMP modes calculated from the delay times by
$v=L/t_{0}$ are presented in Figs.~\ref{fig:vel}(a,b).
At $\nu<3$ the velocity of the fast mode decreases
monotonously achieving a value of $150$\,km/s at filling
factor $\nu=2$. The velocity of the slower mode decreases
proportional to $1/(B-B_0)$ with $B_0=2.48$\,T. For
$B>2.80$\,T it cannot be observed any more in our
measurement window of $800$\,ns.

The velocities of (I)EMP modes can be phenomenologically
described by \mbox{$v\approx (e\Delta n/B)(\alpha/\epsilon_0
\epsilon_r)$} where $\Delta n$ is the difference in the
charge density on both sides of the propagation region of
each mode and $\alpha\le 1$ is a parameter which depends on
the shape of the density profile \cite{sommerfeld95}. Taking
$a^{(2)}$ as the separating strip between the two modes, the
ratio $\Delta n^{\text{EMP}}/\Delta n^{\text{IEMP}}$ grows
from $2.0$ to $2.73$ when the magnetic field increases from
$2.55$ to $2.80$\,T. This change cannot explain the
experimentally observed larger difference in the velocities:
$v^{\text{EMP}}/v^{\text{IEMP}}$ varies from $3.3$ to $17.5$
in this range of magnetic fields. Therefore, different
shapes of density profiles $\alpha^{\text{EMP}}$ and
$\alpha^{\text{IEMP}}$, and their significant change with
magnetic field have to be taken into account. Such magnetic
field dependence is in accordance with the EC picture of the
quantum Hall effect, where the compressible regions of
varying charge density are expected to get broader with
increasing field \cite{chkl92}.

To estimate the spatial extension of the density profile for
each mode we use a local capacity approximation developed in
Ref.~\onlinecite{zhit95}. It connects the group velocity
$v_g$ of the EMP wave packet with the width $l$ of the
propagation region and the distance $d$ between the 2DES and
the metallic top gate:
\begin{equation}
    v_g=\frac{\partial \omega}{\partial
    k}=\frac{e^2\Delta\nu}{h}\frac{d}{\epsilon_0 \epsilon_r
    l},\ \ \ d \ll l \label{eqn:v}
\end{equation}
where $\Delta\nu$ is the difference between the filling
factors of the restricting isolating regions to the left and
to the right of the propagation strip. For the EMP mode
\mbox{$\Delta\nu=g-0=2$}\,, for the IEMP we take
\mbox{$\Delta\nu=\nu-g$}. The magnetic field dependence of
the (I)EMP velocities manifests itself in the increase of
the widths $l_1$ and $l_2$ according to Eq.~(\ref{eqn:v}).
The calculated widths of the propagation regions associated
with the EMP and the IEMP modes are shown in
Figs.~\ref{fig:vel}(c,d). The obtained values exceed
substantially the distance to the top gate for both modes
justifying the local capacity approach. The width of the
outer compressible strip can be estimated from theory
\cite{chkl92}. We obtain $l_1=190$\,nm at $B=2.5$\,T
increasing to $l_1=277$\,nm at $B=2.7$\,T. This agrees quite
reasonable with the experimental result for the EMP mode
[Fig.~\ref{fig:vel}(c)]. The observed plateau-like structure
at $\nu>3$, however, cannot be appropriately accounted for
by a theory, where the widths of compressible strips depend
monotonously on the magnetic field.

In our experiment we observe a distinct transition from
coupled to decoupled transport regimes at $\nu\leq3$, when
the IEMP mode shows up. The decoupling of the innermost edge
state is known to be asymmetric with respect to the quantum
Hall plateau. The authors of Ref.~\onlinecite{alphenaar90}
proposed a hybridization of the current-carrying edge states
with localized states of the same Landau level when this
level approaches the Fermi energy. Then, the wave function
of the innermost EC increases its extension and reduces at
the same time its amplitude near the edge. This means, on
the one hand, a decrease in $\alpha^{\text{IEMP}}$ and, on
the other hand, an increase in the effective width of the
incompressible strip $a$ which enables the observed
decoupling. In addition, the spin gap governing $\nu=3$ can
be strongly enhanced by interaction effects influencing the
actual charge distribution.

The obtained overall sizes of the edge structure $(l_1+l_2)$
are of the same order of magnitude as in previous estimates
from, e.~g., capacitance measurements \cite{takaoka94} and
scanned potential microscopy \cite{mccormick99}. From our
analysis, however, we are able to resolve the internal
spatial structure of edge states in a range of magnetic
fields where their individual contributions can be
unambiguously identified.

The decoupled fundamental modes of EMP's assigned to
different Landau levels have been recently theoretically
treated in Ref.~\onlinecite{balev}. It was shown that both
modes are nearly undamped and propagate with sufficiently
different group velocities. The chosen confinement
potential, however, was sufficiently steep to neglect the
formation of incompressible and compressible strips. Our
observed changes in the velocities which are especially
strong for the IEMP mode suggest that the spatial extent of
the charge density profile connected with each mode must
also be taken into account. Here, more theoretical work is
necessary.

In summary, we employed selective detection of edge states
for time-resolved current measurements under conditions of
strongly screened Coulomb interactions. We observed
decoupled EMP and IEMP modes propagating along neighboring
compressible strips in the vicinity of filling factor $3$.

The authors wish to thank R.~Winkler for a critical reading
of the manuscript and acknowledge financial support by DFG
priority programm ``Quantum Hall Systems'' and the BMBF.



\end{document}